\documentstyle[prl,aps,epsf]{revtex}   
\begin{document}   
\tightenlines
\title{Pairs of charged heavy-leptons from an  SU(3)$_{L}\otimes$U(1)$_{N}$ model at CERN LHC}   
\author{J.\ E.\ Cieza Montalvo}   
\address{Instituto de F\'{\i}sica, Universidade do Estado do Rio de Janeiro, Rua S\~ao Francisco Xavier 524, 20559-900 Rio de Janeiro, RJ, Brazil}   
\author{M. D. Tonasse}   
\address{Instituto Tecnol\'ogico de Aeron\'autica, Centro T\'ecnico  Aeroespacial, Pra\c ca Marechal do Ar Eduardo Gomes 50, 12228-901 S\~ao Jos\'e dos Campos, SP, Brazil}   
\date{\today}   
\maketitle     
\begin{abstract}   
One of the versions of the SU(3)$_L\otimes$U(1)$_N$ electroweak model predicts  charged heavy-leptons which do not belong to any class of heavy-leptons proposed up to now. We investigate the production and signatures of pairs of these heavy-leptons {\it via} the Drell-Yan process and the gluon-gluon fusion at the CERN Large Hadron Collider (LHC). As an example we study the decay of the exotic leptons into another ones. We see that the lifetime of the exotic leptons can be short.
\end{abstract}     
  

\section{Introduction}    

Measurements of the total width of the $Z$ neutral gauge boson of the  standard model at CERN and 
SLAC colliders provide an undeniable evidence for only three neutrino flavors \cite{Cea98}. 
 CP-violation mechanism and big-bang nucleosynthesis, in the framework of the standard model, reinforce the idea  that in the Nature  there are only three families of fermions.\par 
However, since the standard model must be only a low energy effective electroweak theory, there are several motivations to consider the possibility of additional quarks and leptons. For example, grand unified theories such as SO(10) and E(6) incorporate new fermions naturally \cite{MO91}. In some extensions of the standard model, anomalies are canceled by introducing additional fermions \cite{FG87} and this new particles can play a role in CP-violation schemes \cite{FK91}. Heavy-leptons phenomenology are widely studied in many extended electroweak models, such as supersymmetric \cite{cie1}, grand unified theories \cite{la}, technicolor \cite{di}, superstring-inspired models \cite{e6}, mirror  fermions \cite{maa}, etc. All these models predict 
the existence of new  particles with masses around of the scale of $1$ TeV and they consider the possible existence of new generations of fermions.\par 
The standard electroweak model provides a very satisfactory description of most elementary particle phenomena up to the presently available energies. However, there are some unsatisfactory features as the family number and their complex pattern of masses and mixing angles, which are not predicted by the model. In addition, the recently emerged  experimental data on the muon anomalous magnetic moment  \cite{Bea01}, solar and atmospheric neutrinos \cite{Aea99} strongly suggest physics beyond the standard model. Searches on new fermions can reveals some directions for the solution of these problems.\par 
In this paper we addresses to a new class of heavy-fermions, {\it i.e.}, the heavy-leptons which can be produced by the strong and electroweak processes which emerge from a model based on the SU(3)$_{C}\otimes$SU(3)$_{L}\otimes$U(1)$_{N}$ (3-3-1 for short) semi simple symmetry group \cite{PT93}. In this model the new leptons do not require new generations, as occur in the most of the heavy-lepton models \cite{FH99}. This ones is a chiral electroweak model whose left-handed charged heavy-leptons, which we denote by $P_a$ $=$ $E$, $M$ and $T$ $\left(a = 1, 2, 3\right)$, together with the associated ordinary charged leptons and its respective neutrinos, are accommodate in SU(3)$_L$ triplets. We study the production of these charged heavy-lepton pairs in hadronic collisions at the CERN Large Hadron Collider (LHC). This process will be studied through the well known Drell-Yan mechanism, {\it i. e.}, the quark-antiquark fusion $q \bar q \rightarrow P^-P^+$ (Fig. 1a) and the gluon-gluon fusion $gg \rightarrow P^{-} P^{+}$ (Fig. 1b).\par
We remember at first that the heavy-leptons appearing in the literature up to now can be classified in four kinds \cite{Cea98}: (a) Sequential leptons, in that the new leptons are associated with new neutrinos, forming new SU(2)$_L$ doublets. This new leptons and its neutrinos have the same leptonic numbers which are conserved in all interactions; (b) paraleptons, in that the new leptons have the same leptonic numbers of the associated ordinary charged leptons of opposite charge; (c) ortholeptons, in that the new leptons have the same leptonic numbers of the associated ordinary charged leptons of the same charge and (d) long-lived penetrating particles. As we mentioned above there are very motivations to search heavy-leptons in electroweak extensions of the standard model. However, in 3-3-1 model we have in addition two particular motivations. Firstly, as we will see in the next section, the 3-3-1 heavy-leptons do not belong to any of these types of new leptons. Consequently, the existing experimental bounds on heavy-lepton parameters do not apply to them. Secondly, there is nothing on the 3-3-1 heavy-lepton phenomenology up to now.\par 
\begin{figure}[h]
\begin{center}  
\epsfxsize=4in  
\epsfysize=5 true cm  
\centerline{\epsfbox{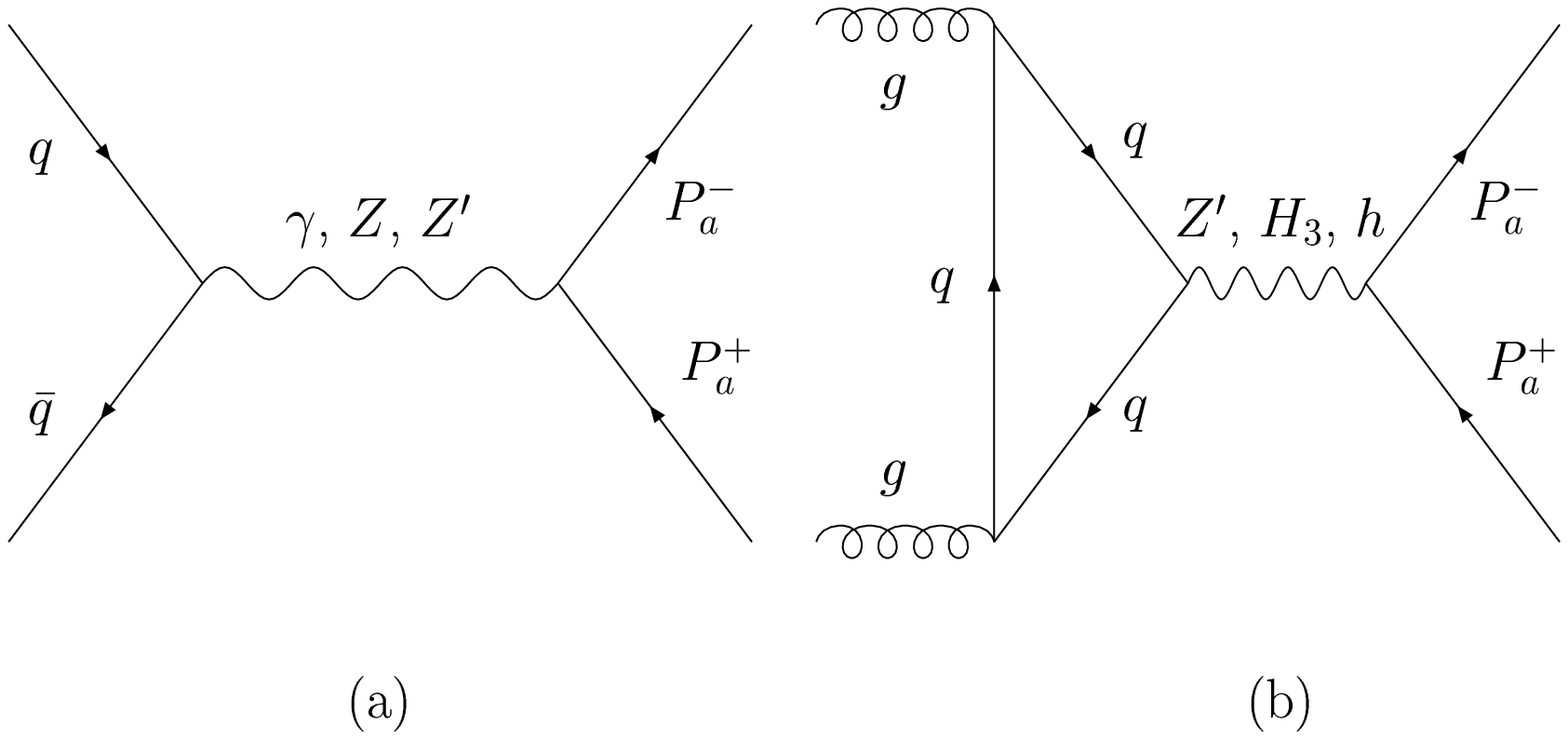}}  
\end{center}  
\caption[]{Feynman diagrams for production of charged heavy-lepton pairs {\it via} (a) Drell-Yan process and (b) gluon-gluon fusion.}
\label{fig:1}\end{figure}
The outline of this paper is the following. In Sec. II we present the relevant features of the model. The production of a pair of 3-3-1 heavy-leptons in $pp$ colliders are discussed in Sec. \ref{secIII} and in Sec. IV we summarize our results and conclusions.     

\section{The 3-3-1 heavy-lepton model}   
\label{secII}    

In 3-3-1 model the strong and electroweak interactions are described by a gauge theory based on the SU(3)$_C\otimes$SU(3)$_L\otimes$U(1)$_N$ semi simple symmetry group. In its original version new leptons are not required, since the lepton representation content of each SU(3)$_L$ triplet consists of one charged  lepton, its charge conjugated counterpart and the associated neutrino field 
\cite{PP92,FR92}. However, the version of the model which we work here differs from the original one, in that the charge conjugated lepton fields of the original model are replaced by heavy-leptons in the SU(3)$_L$ triplets\cite{PT93}. The most interesting feature of this class of models is related with the anomaly cancellations, which is implemented only when the three fermion families are considered together and not family by family as in the standard model. This implies that the number of families must be a multiple of the color number and, consequently, the 3-3-1 model suggests a route towards the response of the flavor question \cite{PP92}. The model has also a great phenomenological interest since the related new physics can be expected in a scale near of the Fermi one \cite{CQ99}.\par
Let us summarize the most relevant points of the model (for details see Refs. 
\cite{PT93,PP92,FR92}). In its heavy-lepton version the left-handed leptons and quarks transform under the SU(3)$_L$ gauge group as the triplets   
\begin{mathletters}  
\begin{equation}  
\psi_{aL} = \left(\begin{array}{c}  \nu_{\ell_a} \\  \ell_a^\prime \\  P^\prime_a 
\end{array}\right)_L \sim \left({\bf 3}, 0\right), \quad Q_{1L} = \left(\begin{array}{c} 
u^\prime_1 \\ d^\prime_1 \\  J_1 \end{array}\right)_L \sim \left({\bf 3}, \frac{2}{3}\right), 
\quad Q_{\alpha L} = \left(\begin{array}{c} J^\prime_\alpha \\  u^\prime_\alpha \\   
d^\prime_\alpha \end{array}\right)_L \sim \left({\bf 3}^*, -\frac{1}{3}\right),  
\label{cont} 
\end{equation}\noindent  
where $P^\prime_a$ $=$ $E^\prime$, $M^\prime$, $T^\prime$ are the new leptons, $\ell^\prime_a$ 
$=$ $e^\prime$, $\mu^\prime$, $\tau^\prime$ and $\alpha$ = 2, 3. The $J_1$ exotic quark carries 
$5/3$ units of elementary electric charge while $J_2$ and $J_3$ carry $-$4/3 each. In Eqs. (\ref{quark}) the numbers 0, 2/3 and $-$1/3 are the U(1)$_N$ charges. Except the neutrino fields, which we are considering massless here, each left-handed fermion has its right-handed counterpart transforming as a singlet in the presence of the SU(3)$_L$ group, {\it i. e.},    
\begin{eqnarray}  
\ell'_R \sim \left({\bf 1}, -1\right), & \qquad & P'_R \sim \left({\bf 1}, 1\right), \qquad U'_R 
\sim \left({\bf 1}, 2/3\right), \\  D'_R \sim \left({\bf 1}, -1/3\right), & \qquad & J_{1R} \sim 
\left({\bf 1}, 5/3\right), \qquad J_{2,3R}' \sim \left({\bf 1}, -4/3\right).   
\end{eqnarray}
\label{quark}
\end{mathletters}\noindent 
We are defining $U = u, c, t$ and $D = d, s, b$. In order to avoid anomalies, one of the quark families must transforms in a different way with respect to the two others. In Eqs. (\ref{quark}) all the primed fields are linear combinations of the mass eigenstates. The charge operator is defined by   
\begin{equation}  
\frac{Q}{e} = \frac{1}{2}\left(\lambda_3 - \sqrt{3}\lambda_8\right) + N,  
\label{op} 
\end{equation}  
where the $\lambda$'s are the usual Gell-Mann matrices. We notice, however, that since $Q_{\alpha L}$ in Eqs. (\ref{cont}) are in antitriplet representation of SU(3)$_L$, the 
antitriplet representation of the Gell-Mann matrices must be used in Eq. (\ref{op}) in order to get the correct electric charge for the quarks of the second and third generations.\par    
The three Higgs scalar triplets
\begin{equation}
\eta = \left(\begin{array}{c} \eta^0 \\  \eta_1^- \\  \eta_2^+ \end{array}\right) \sim \left({\bf 3}, 0\right), \quad \rho = \left(\begin{array}{c} \rho^+ \\  \rho^0 \\  \rho^{++}
\end{array}\right) \sim \left({\bf 3}, 1\right), \quad \chi =
\left(\begin{array}{c} \chi^- \\
\chi^{--} \\ \chi^0 \end{array}\right) \sim \left({\bf 3}, -1\right),
\label{higgs}\end{equation}
generate the fermion and gauge boson masses in the model. The neutral scalar fields develop the vacuum expectation values (VEVs)  $\langle\eta^0\rangle = v_\eta$, $\langle\rho^0\rangle = v_\rho$ and  $\langle\chi^0\rangle = v_\chi$, with $v_\eta^2 + v_\rho^2 = v_W^2 = (246 \mbox{ GeV})^2$. The pattern of symmetry breaking is   
\[ \mbox{SU(3)}_L \otimes\mbox{U(1)}_N \stackrel{\langle\chi\rangle}{\longmapsto}\mbox{SU(2)}_L\otimes\mbox{U(1)}_Y 
\stackrel{\langle\eta, \rho\rangle}{\longmapsto}\mbox{U(1)}_{\rm em}\] 
and so, we can expect $v_\chi \gg v_\eta, v_\rho$. The $\eta$ and $\rho$ scalar triplets give masses to the ordinary fermions and gauge bosons, while the $\chi$ scalar triplet gives masses to the new fermions and new gauge bosons. The most general, gauge invariant and renormalizable Higgs potential is 
\begin{eqnarray}
V\left(\eta, \rho, \chi\right) & = & \mu_1^2\eta^\dagger\eta + \mu_2^2\rho^\dagger\rho + \mu_3^2\chi^\dagger\chi + \lambda_1\left(\eta^\dagger\eta\right)^2 + \lambda_2\left(\rho^\dagger\rho\right)^2 + \lambda_3\left(\chi^\dagger\chi\right)^2 + \cr
&& \left(\eta^\dagger\eta\right)\left[\lambda_4\left(\rho^\dagger\rho\right) + \lambda_5\left(\chi^\dagger\chi\right)\right] + \lambda_6\left(\rho^\dagger\rho\right)\left(\chi^\dagger\chi\right) + \lambda_7\left(\rho^\dagger\eta\right)\left(\eta^\dagger\rho\right) + \cr
&& \lambda_8\left(\chi^\dagger\eta\right)\left(\eta^\dagger\chi\right) + \lambda_9\left(\rho^\dagger\chi\right)\left(\chi^\dagger\rho\right) + \lambda_{10}\left(\eta^\dagger\rho\right)\left(\eta^\dagger\chi\right) + \cr
&& \frac{1}{2}\left(f\epsilon^{ijk}\eta_i\rho_j\chi_k + {\mbox{H. c.}}\right).
\label{pot}\end{eqnarray} 
Here $f$ is a constant with dimension of mass and the $\lambda_i$, $\left(i = 1, \dots, 10\right)$ are adimensional constants with $\lambda_3 < 0$ from the positivity of the scalar masses. The term proportional to $\lambda_{10}$ violates lepto-barionic number and so, it was not considered in the analysis of the Ref. \cite{TO96} (another analysis of the 3-3-1 scalar sector are given in Ref. \cite{AK00} and references cited therein). We can notice that this term contributes to the mass matrices of the charged scalar fields, but not to the neutral ones.  However, can be checked that in the approximation $v_\chi \gg v_\eta, v_\rho$ we can still work with the masses and eigenstates given in Ref. \cite{TO96}. Here this term is important to the decay of the ligthest exotic fermion. Therefore, we are keeping it in the Higgs potential.\par
Symmetry breaking is initiated when the scalar neutral fields are shifted as $\varphi = v_\varphi + \xi_\varphi + i\zeta_\varphi$, with $\varphi$ $=$  $\eta^0$, $\rho^0$, $\chi^0$. Thus, the physical neutral scalar eigenstates  $H^0_1$, $H^0_2$, $H^0_3$ and $h^0$ are related to the shifted fields as    
\begin{mathletters}\begin{equation}  
\left(\begin{array}{c} \xi_\eta \\  \xi_\rho \end{array}\right) \approx
\frac{1}{v_W}\left(\begin{array}{cc} v_\eta & v_\rho \\  v_\rho & -v_\eta 
\end{array}\right)\left(\begin{array}{c} H^0_1 \\  H^0_2 \end{array}\right), \qquad \xi_\chi \approx H^0_3, \qquad \zeta_\chi  \approx h^0, 
\label{eign}\end{equation}
and in the charged scalar sector we have
\begin{equation}
\eta^+_1 \approx \frac{v_\rho}{v_W}H^+_1, \qquad \rho^+ \approx \frac{v_\eta}{v_W}H_2^+, \qquad \chi^{++} \approx \frac{v_\rho}{v_\chi}H^{++},
\label{eigc}\end{equation}\label{eig}\end{mathletters}
with the condition that $v_\chi \gg v_\eta, v_\rho$ \cite{TO96}. From Eqs. (\ref{quark}) to (\ref{eig}) we can notice that the ordinary quarks couple only through $H^0_1$ and $H^0_2$ and the heavy-leptons and quarks couple only through $H^0_3$ and $h^0$. So, it is easy to see that in this regime there will be no contribution of the Higgs bosons to the diagrams analogous to one of the Figs. \ref{fig:1}. \par 
Due the transformation properties of the fermion and Higgs fields under SU(3)$_L$ [see Eqs. (\ref{quark}) and (\ref{higgs})] the Yukawa interactions in the model are    
\begin{mathletters} 
\begin{eqnarray}
{\cal L}_\ell^Y & = & -G_{ab}\bar\psi_{aL}\ell^\prime_{bR}\rho - G^\prime_{ab}\bar\psi_{aL}^\prime P^\prime_{bR}\chi + \mbox{H. c.}, 
\label{yl} \\ 
{\cal L}_q^Y & = & \sum_a\left[\bar Q_{1L}\left(G_{1a}U^\prime_{a R}\eta + \tilde G_{1a}D^\prime_{a R}\rho\right) + \sum_\alpha\bar  Q_{\alpha L}\left(F_{\alpha a}U^\prime_{aR}\rho^* +  \tilde F_{\alpha a}D^\prime_{aR}\eta^*\right)\right] + \cr && 
\sum_{\alpha\beta}F^J_{\alpha\beta}\bar Q_{\alpha L}J^\prime_{\beta R}\chi^* + G^J\bar  Q_{1L}J_{1R}\chi +  \mbox{H. c.}  
\label{yq}
\end{eqnarray}\label{weak}\end{mathletters}\noindent
The $G$'s, $\tilde G$'s, $F$'s and $\tilde F$'s are Yukawa coupling constants with $a, b = 1, 2, 3$ and $\alpha, \beta = 2, 3$. The interaction eigenstates which appear in Eqs. (\ref{weak}) can be transformed in the corresponding physical eigenstates by appropriated rotations. However, since the cross section calculation imply summation on flavors (see Sec. \ref{secIII}) and  the rotation matrix must be unitary, the mixing parameters have not essential effect for our purpose here. So, thereafter we suppress the primes notation for the interactions eigenstates.\par    
The gauge bosons consist of an octet $W^i_\mu$ $\left(i = 1, \dots, 8\right)$ associated with SU(3)$_L$ and a singlet $B_\mu$ associated with U(1)$_N$. The covariant derivatives are
\begin{equation}
{\cal D}_\mu\varphi_a = \partial_\mu\varphi_a+ i\frac{g}{2}\left(\vec W_\mu.\vec\lambda\right)^b_a\varphi_b + ig^\prime N_\varphi\varphi_aB_\mu,
\end{equation}
where $\varphi = \eta, \rho, \chi$. The model predicts single charged $\left(V^\pm\right)$, double charged $\left(U^{\pm\pm}\right)$ vector bileptons and a new neutral gauge boson $\left(Z^\prime\right)$ in adition to the charged standadrd gauge bosons $W^\pm$ and the neutral standard $Z$. We take the physical eigenstates of the neutral gauge bosons from Refs. \cite{PP92},   
\begin{mathletters}  
\begin{eqnarray}  
A_{\mu} & = & \sqrt{\frac{3}{f\left(t_W\right)}}\left[\left(W_\mu^3 -  \sqrt{3}W^8_\mu\right)t_W + B_\mu\right], \\  
Z_\mu & \simeq & -\sqrt{\frac{3}{f\left(t_W\right)}}\left(\sqrt{1 +  3t_W^2}W^3_\mu + 
\sqrt\frac{3}{1 + 3t_W^2}t_W^2W_\mu^8 - \frac{1}{\sqrt{1 +  3t_W^2}}B_\mu\right), \\   
Z^\prime_\mu & \simeq & \frac{1}{\sqrt{1 + 3t_W^2}}\left(W^8_\mu +  \sqrt{3}t_WB_\mu\right).  
\end{eqnarray}\label{azz}\end{mathletters}\noindent
Here the $Z$ and $Z^\prime$ eigenstates are valid in the approximation $v_\chi \gg v_\eta, v_\rho$. We can see that in this approximation $Z^\prime$ does not interacts with the standard $W^\pm$ gauge bosons. In Eqs. (\ref{azz})    
\begin{equation}  
t_W^2 = \frac{\sin^2{\theta_W}}{1 - 4\sin^2{\theta_W}}  
\label{tw}
\end{equation} 
and $f^2\left(t_W\right) = 3\left(1 + 4t_W^2\right)$. We observe that the $t_W$ parameter in Eq. (\ref{tw}) has a Landau pole which imposes $\sin^2{\theta_W} < 1/4$. It is another good feature of the model, since evolving the Weinberg angle $\sin^2{\theta_W}$ to high values it is possible to find an upper bound to the masses of the new charged gauge bosons \cite{FR92,JJ97}. It does not occur in the popular extensions of the standard model. Another consequence of this pole is an enhancement of the couplings of the $Z'$ to quarks and to the leptons.\par
\begin{figure}[h]
\begin{center}  
\epsfxsize=3in  
\epsfysize=3 true cm  
\centerline{\epsfbox{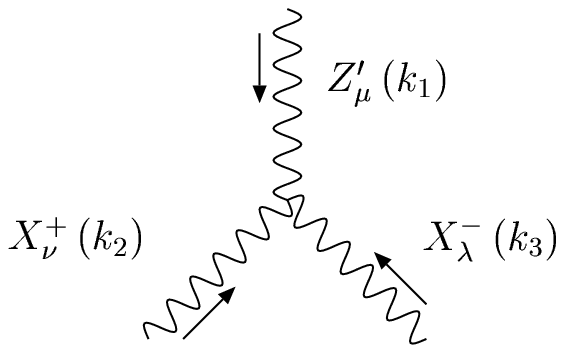}}  
\end{center}  
\caption[]{Trilinear interaction among $Z^\prime$ and the exotic charged  gauge bosons $\left(X^\pm_\mu = V^\pm_\mu, U^{\pm\pm}_\mu\right)$.}
\label{fig:2}\end{figure}

    
The trilinear interactions of the $Z^\prime$ with the $V^\pm$ and $U^{\pm\pm}$ bileptons has strenght
\begin{equation}
{\cal V}_{\lambda\mu\nu} = -i\frac{g}{2}\sqrt{\frac{3}{1 + 3t_W^2}}\left[\left(k_1 - k_2\right)_\lambda g_{\mu\nu} + \left(k_2 - k_3\right)_\mu g_{\nu\lambda} + \left(k_3 - k_1\right)_\nu g_{\lambda\mu}\right]
\end{equation}
with the quadrimoments defined in Fig. \ref{fig:2}. The relevant neutral vector current interactions are    
\begin{mathletters}
\begin{eqnarray}
{\cal L}_{AP} & = & -e\bar P_a\gamma^\mu P_aA_\mu, \\
{\cal L}_{ZP} & = & -g\sin{\theta_W}\tan{\theta_W}\bar P_a\gamma^\mu P_aZ_\mu,
\label{lagrb}\\
{\cal L}_{Z^\prime P} & = &  -\frac{g\tan{\theta_W}}{2\sqrt{3}t_W}\bar P_a\gamma^\mu\left[3t_W^2 - 1 + \left(3t_W^2 + 1\right)\gamma_5\right]P_aZ^\prime_\mu,
\label{lagrc}\\
{\cal L}_{Zq} & = & -\frac{g}{4\cos{\theta_W}}\sum_a\bar q_a\gamma^\mu\left(v^a + a^a\gamma^5\right)q_aZ_\mu,
\label{lagrd}\\
{\cal L}_{Z^\prime q} & = & -\frac{g}{4\cos{\theta_W}}\sum_a\bar q_a\gamma^\mu\left(v^{\prime a} 
+ a^{\prime a}\gamma^5\right)q_aZ^\prime_\mu,
\label{lagre}
\end{eqnarray}
\label{lagr}
\end{mathletters}\noindent 
where $q_a$ is any quark  \cite{PT93,PP92}. The coefficients in Eqs. (\ref{lagrd}) and (\ref{lagre}) are   
\begin{figure}[h]
\begin{center}  
\epsfxsize=2in  
\epsfysize=4 true cm  
\centerline{\epsfbox{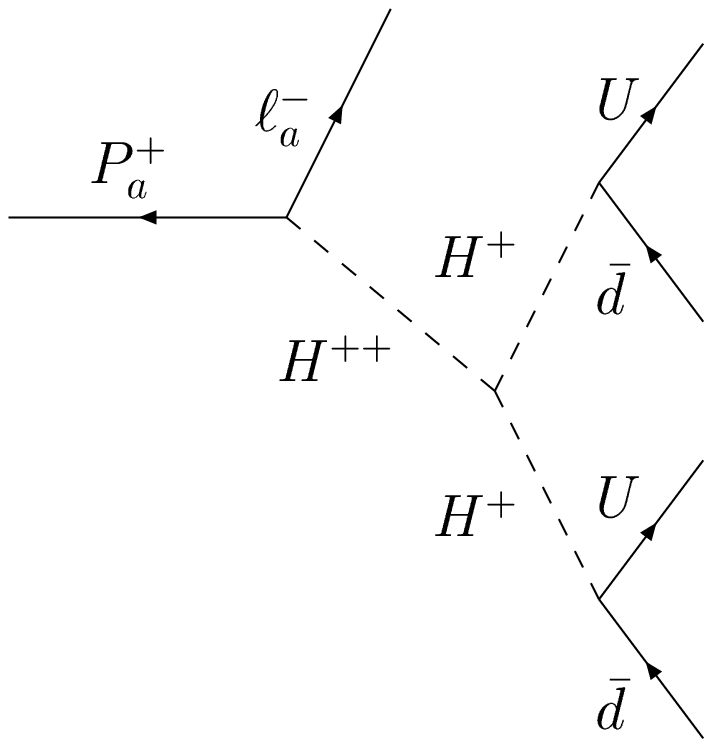}}  
\end{center}  
\caption[]{Example of how an exotic 3-3-1 particle (the heavy-leptons $P_a$) can decay in standard particles {\it via} Higgs sector. Here $U = u, c, t$.}
\label{fig:3}\end{figure}
\begin{mathletters}  
\begin{eqnarray}  
v^U & = & \frac{3 + 4t_W^2}{3\left(1 + 4t_W^2\right)}, \quad v^D = -\frac{3 + 8t_W^2}{3\left(1 + 
4t_W^2\right)}, \quad -a^U = a^D = 1, \quad  v^{\prime u} = -\frac{1 + 
8t_W^2}{f\left(t_W\right)}, \\  v^{\prime c} & = & v^{\prime t} = \frac{1 - 
2t_W^2}{f\left(t_W\right)}, \quad v^{\prime d} = -\frac{1 + 2t_W^2}{f\left(t_W\right)}, \quad 
v^{\prime s} =  v^{\prime b} = \frac{f\left(t_W\right)}{\sqrt(3)}, \quad a^{\prime u} =  
\frac{1}{f\left(t_W\right)}, \\ 
a^{\prime c} & = & a^{\prime t} = -\frac{1 + 6t_W^2}{f\left(t_W\right)}, \quad a^{\prime d} = 
-a^{\prime c}, \quad a^{\prime s} = a^{\prime b} = -a^{\prime u} \quad v^{\prime J_1} = 
\frac{2\left(1 - 7t_W^2\right)}{f\left(t_W\right)}, \\ v^{\prime J_2} & = & v^{\prime J_3} = 
-\frac{2\left(1 - 5t_W^2\right)}{f\left(t_W\right)}, \quad
a^{\prime J_2} = a^{\prime J_2} = a^{\prime J_2} = a^{\prime J_1} = -\frac{2\left(1 + 
3t_W^2\right)}{f\left(t_W\right)}
\end{eqnarray} 
\label{coef}
\end{mathletters} 
\begin{figure}[h]
\begin{center}  
\epsfxsize=4 true in  
\epsfysize=5 true cm  
\centerline{\epsfbox{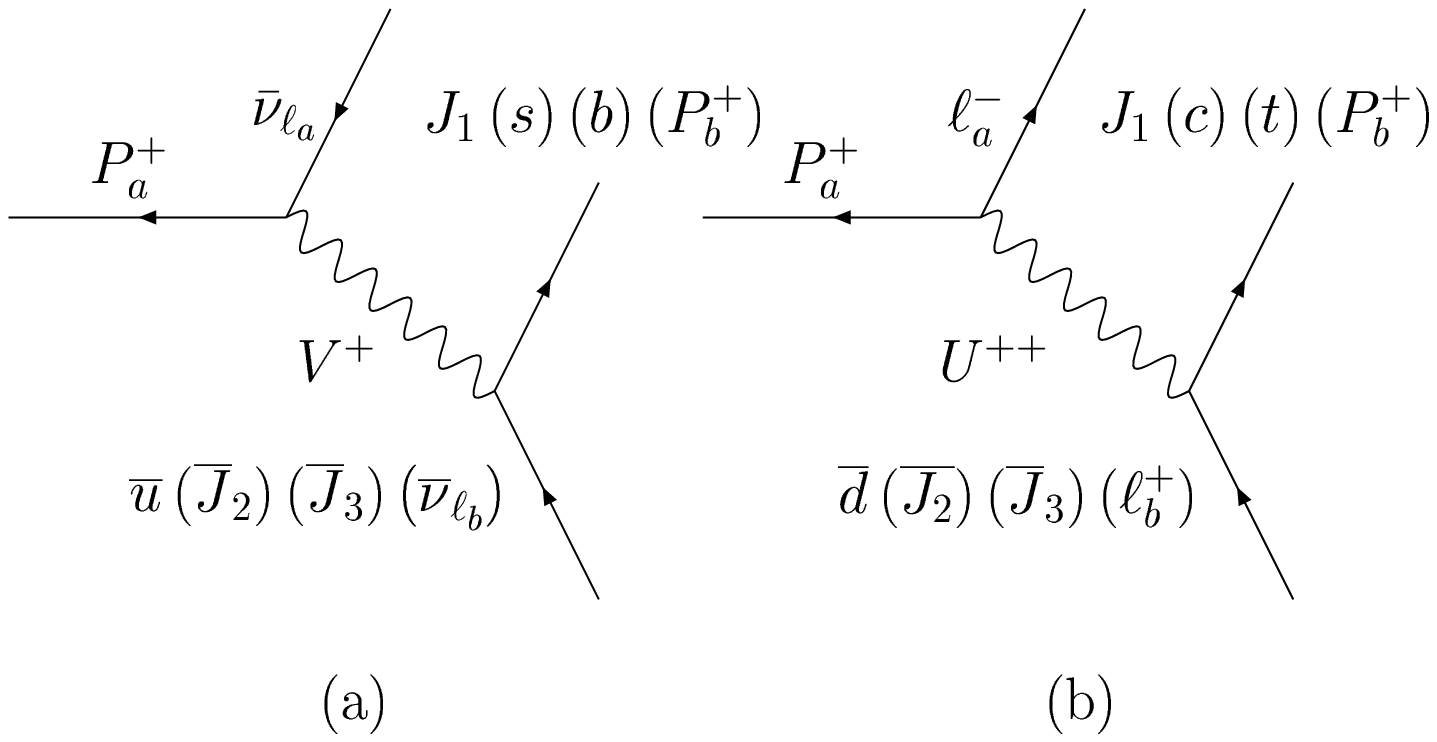}}  
\end{center}  
\caption[]{Decay channels for the 3-3-1 heavy-leptons (a) {\it via} $V^+$ and (b) {\it via} $U^{++}$.}
\label{fig:4}\end{figure}
As we comment in Sec. I and by inspection of Eqs. (\ref{quark}), (\ref{lagrb}) and (\ref{lagrc}), we conclude that the heavy-leptons $P_a$ belong to another  class of exotic particles differently of the heavy-lepton classes usually considered in the literature. Thus, the  present experimental limits 
do not apply directly to them \cite{Cea98} (see also Ref. \cite{PT93}). Therefore, the 3-3-1 heavy-leptons phenomenology deserves more detailed studies.\par    
The charged lepton current interactions are   
\begin{mathletters}  
\begin{equation}  
{\cal L}^{CC}_\ell = -\frac{g}{\sqrt{2}}\sum_a\left(\overline{\ell'}_{aL}\gamma^\mu\nu_{aL} 
W^-_{\mu} + \overline{P'}_{aL}\gamma^\mu\nu_{bL}V^+_\mu +  \overline{\ell'}_{aL}\gamma^\mu 
P'_{bL}U^{--}\right) + {\mbox{H. c.}}  
\end{equation} 
For the first and second generations of quark we have    
\begin{eqnarray}  
{\cal L}^{CC}_{Q_1} & = & -\frac{g}{\sqrt{2}}\left(\overline{u'}_L\gamma^\mu  d'_LW^-_\mu + 
\overline J_{1L}\gamma^\mu u'_LV^+_\mu + \overline{d'}_L \gamma^\mu J_{1L}U^{--}\right) + 
{\mbox{H. c.}}, \\ {\cal L}^{CC}_{Q_2} & = & -\frac{g}{\sqrt{2}}\left(\overline{c'}_L\gamma^\mu  
d'_LW^-_\mu - \overline{s'}_L\gamma^\mu J'_{2L}V^+_\mu +  \overline{c'}_L\gamma^\mu 
J'_{2L}U^{--}\right) + {\mbox{H. c.}},  
\end{eqnarray}
\label{llq}
\end{mathletters}\noindent
while the charged current interactions for the quarks of the third generation are  obtained from those 
of the second generation replacing $c \to t$, $s \to b$ and $J_2 \to J_3$.\par 
Also, the main decay modes of the new leptons are among the exotic leptons themselves such as $T^{+} \rightarrow 
E^{+} \nu_{e} \bar{\nu_{\tau}}$, considering that $M_{T} > M_{E}$. The heavy-leptons can decay also {\it via} single or double charged bileptons in a standard lepton, a standard quark and an exotic quark (see Fig. \ref{fig:3}). Detailed analysis of the 3-3-1 heavy-lepton decays will be given elsewhere \cite{CT00}. The lightest exotic fermion of the model can decay in the ordinary particles {\it via} the Higgs sector. Looking for the term proportional to $\Lambda_{10}$ in the Higgs potential (\ref{pot}), we see that after the symmetry breaking this term leads to the couplings $\lambda_{10}\eta^+_1\chi^{--}\left(v_\eta\rho^+ + v_\rho\eta_1^+\right)$, with the charged scalar fields given by Eqs. (\ref{eigc}). From Eq. (\ref{yl}) it is easy to see that the scalar $\chi^{--}$ couples the exotic leptons $P_a$ to ordinary ones $\ell_a$ and from Eq. (\ref{yq}) the  $\eta_1^+$ couples ordinary quarks of $-$1/3 charge to ones of 2/3 charge. Therefore, the decay of the exotic leptons $P_a$ can proceeds as in the diagram of the Fig. \ref{fig:3}.\par
Inspection of Eq. (\ref{yq}) tell us also that the exotic quarks can decay in a similar form as if in the Fig. \ref{fig:4}, where the ordinary leptons $\ell_a$ would be  replaced by the appropriate standard quarks. Therefore, the lightest exotic fermion can be not stable.\par

\section{Cross section production}   
\label{secIII}    

To calculate the cross-section, we begin suppressing the family indices of the Sec. \ref{secII}, 
{\it i. e.}, we assume $P$ $=$ $E$, $M$, $T$; $\ell$ $=$ $e$, $\mu$, $\tau$ and $J$ $=$ $J_1$, $J_2$, $J_3$. We will study  
the mechanism of the  Drell-Yan production of pair of heavy-leptons, that is, we analyze the   
process $pp  \rightarrow q \bar{q} \rightarrow P^-P^+$ (Fig. \ref{fig:1}a). This process take place through the exchange of the bosons $Z$, $Z^\prime$ and $\gamma$ in the $s$ channel.    
Using the interaction Lagrangians of the Eqs. (\ref{lagr}) and the parameters of the Eqs. (\ref{coef}), we evaluate the differential cross section for the subprocess $q \bar{q} 
\rightarrow P^-P^+$ obtaining
\begin{eqnarray}
\left(\frac{d\sigma}{d\cos\theta}\right)_{P^+P^-} & = & \frac{\beta\alpha^2\pi}{N_cs^2}\left\{\frac{e^2_q}{s}\left[2sM_P^2 + \left(m_P^2 - t^2\right)^2 + \left(m_P^2 - u^2\right)^2\right] \right.\cr
&&\left. - \frac{e_q}{2\sin^2\theta_W\cos^2\theta_W\left(s - M^2_{Z,Z'} + iM_{Z,Z'}\Gamma_{Z,Z'}\right)} \right.\cr
&&\left. \times\left[2sM_P^2{g'_V}^{PP}g^q_V + {g'_V}^{PP}g^q_V\left[\left(M_P^2 - t^2\right)^2 + \left(M_P^2 - u^2\right)^2\right] \right.\right.\cr
&&\left.\left. + {g'_A}^{PP}g^q_A\left(\left(M_P^2 - u\right)^2 - \left(M_P^2 - t^2\right)^2\right)\right] \right\}\cr
&& + \frac{\beta\pi\alpha^2}{16 N_c\cos^4\theta_W\sin^4\theta_W} \frac{1}{s\left(s - M_{Z,Z'}^2 + i M_{Z,Z'} \Gamma_{Z,Z'}\right)^2} \cr  
&& \times\left\{\left[\left({g^\prime}_V^{PP}\right)^2 +  
\left({g^\prime}_A^{PP}\right)^2\right]\left[\left(g_V^q\right)^2 + \left(g_A^q\right)^2\right]\left[\left(M_P^2 - u\right)^2 + \left(M_P^2 - t\right)^2\right] \right. \cr 
&& \left. + 2sM_P^2\left[\left({g^\prime}_V^{PP}\right)^2 - \left({g^\prime}_A^{PP}\right)^2\right]\left[\left(g_V^q\right)^2 + \left(g_A^q\right)^2\right]\right. \cr 
&& \left. + 4{g^\prime}_V^{PP}{g^\prime}_A^{PP}g_V^qg_A^q  \left[\left(M_P^2 - u\right)^2 - \left(M_P^2 -t\right)^2\right]\right\}\cr 
&& + \frac{\beta\pi\alpha^2}{8N_c\sin^4\theta_W\cos^4\theta_W s\left(s- M_Z^2 + iM_Z\Gamma_Z\right)\left(s - M_{Z'}^2 + i M_{Z'}\Gamma_{Z'}\right)} \cr
&&\times\left\{2sM_P^2\left(g_V^q + g_A^q\right)\left(g_V^{PP}{g^\prime}_V^{PP} -g_A^{PP}{g^\prime}_A^{PP}\right) + \left(M_{P}^{2}- 
t\right)^2\right.\cr
&&\left.\times\left[\left(\left(g_V^q\right)^2 +  
\left(g_A^q\right)^2\right)(g_V^{PP}{g^\prime}_V^{PP} + g_A^{PP}{g^\prime}_A^{PP}) - 2g_V^q g_A^qg_V^{PP}{g^\prime}_A^{PP} \right.\right.\cr 
&&\left.\left.- 2g_V^qg_A^qg_A^{PP}{g^\prime}_V^{PP}\right] + \left(M_P^2 - u\right)^2\left[\left(\left(g_V^q\right)^2 + \left(g_A^q\right)^2\right)\right.\right.\cr
&&\left.\left.\times\left(g_V^{PP}{g^\prime}_V^{PP} + g_A^{PP}{g^\prime}_A^{PP}\right) + 2g_V^qg_A^qg_V^{PP}{g^\prime}_A^{PP} + 2g_V^qg_A^qg_A^{PP}{g^\prime}_V^{PP}\right]\right\},
\end{eqnarray}
where
\begin{equation}  
g_{V,A}^{PP} = \frac{a_L \pm a_R}{2}, \qquad {g^\prime}_{V,A}^{PP} =  \frac{a^\prime_{L} \pm 
a^\prime_{R}}{2}.  
\end{equation}\noindent 
Here the primes $\left(^\prime\right)$ are for the case when we take a boson $Z'$, $\Gamma_{Z,Z'}$ is 
the total width of the boson Z and $Z'$, $\beta = \sqrt{1 - 4 M_P^{2}/s}$ is the velocity of the 
heavy-lepton in the c.  m. of the process, $e_q$ is the electric charges of the quark $q$, $N_c$ is the  
number of colors, $g^q_{V, A}$ are the standard quark coupling constants, $M_{Z}$ is the mass of 
the $Z$ boson, $\sqrt{s}$ is the center of mass energy of the $q \bar{q}$ system, $t = M_P^{2} - 
\left(1 -\beta \cos \theta\right)s/2$ and $u = M_P^{2} - \left(1 + \beta \cos \theta\right)s/2$, 
where $\theta$ is the angle between the heavy-lepton  and the incident quark in the c. m. frame. 
For $Z^\prime$ boson we take  $M_{Z^\prime} = \left(0.5 - 3\right)$ TeV, since $M_{Z^\prime}$ is  
proportional to VEV $v_\chi$ \cite{PP92,FR92}. For the standard model parameters we assume PDG 
values, {\it i. e.}, $M_Z = 91.19$ GeV, $\sin^2{\theta_W} = 0.2315$ and $M_W = 80.33$ GeV 
\cite{Cea98}.\par
The total width of the $Z^\prime$ boson into exotic leptons, standard  leptons, neutrinos, 
standard and exotic quarks and new vector bosons are   
\begin{equation}  
\Gamma \left(Z' \rightarrow {\rm all}\right)  =  \Gamma_{Z' \rightarrow P^{-}  P^{+}} +  
\Gamma_{Z' \rightarrow \ell^{-} \ell^{+}} + \Gamma_{Z' \rightarrow \nu  \bar{\nu}} + \Gamma_{Z' 
\rightarrow q \bar{q} \left(J \bar{J}\right)} +  2\Gamma_{Z' \rightarrow X^{-} X^{+}}, 
\end{equation} \noindent
where $X^\pm = V^\pm$ or $U^{\pm\pm}$ and we have for everyone   
\begin{mathletters}  
\begin{eqnarray}
\Gamma_{Z' \rightarrow P^{-} P^{+}} & = & \frac{\alpha \sqrt{1- 4 M_{P}^2/  s}}{12 M_{Z'} 
\sin^{2} \theta_{W} \cos^{2} \theta_{W}}  \left  [2 M_{P}^{2}\left(g_{V}^{\prime PP}\right)^{2}  
-  4 M_{P}^{2} \left(g_{A}^{\prime PP}\right)^{2} + M_{Z'}^{2}   \left(g_{V}^{\prime 
PP}\right)^{2} \right.\nonumber \\  
&& \left.+ M_{Z'}^{2}  \left(g_{A}^{\prime PP}\right)^{2} \right],   \\  
\Gamma_{Z' \rightarrow \ell^{-} \ell^{+}} & = & \frac{\alpha M_{Z'}}{12  \sin^{2} \theta_{W}\cos^{2} \theta_{W}} \left[\left(g_{V}^{\ell}\right)^{2} +   \left(g_{A}^{\ell}\right)^{2} 
\right], \qquad \Gamma_{Z' \rightarrow \nu \nu} = \frac{\alpha M_{Z'}}{18h\sin^2{\theta_W}\cos^2{\theta_W}}, \\  
\Gamma_{Z' \rightarrow q \bar{q} \left(J \bar{J}\right)} & = &   \frac{\alpha \sqrt{1- 4 M_{q}^2/s}}{16 M_{Z'} \sin^{2} \theta_{W} \cos^{2} \theta_{W}}\left[2M_{q}^{2}\left(g_{V_{i}}^{qq}\right)^{2}  -  4 M_{q}^{2} \left(g_{A_{i}}^{qq}\right)^{2} + M_{Z'}^{2}   \left(g_V^{qq}\right)^{2} \right.\nonumber \\
 && \left. + M_{Z'}^{2}\left(g_A^{qq}\right)^{2} \right], \\
\Gamma_{Z' \rightarrow Y_1^{-} Y_2^{+}} & = &  \frac{\alpha \sqrt{M_{Z'}^{2} - (M_{Y_{1}}+ M_{Y_{2}})^{2}}  \sqrt{M_{Z'}^{2} - (M_{Y_{1}}- M_{Y_{2}})^{2}}}{8 M_{Z'}^{3} \sin^{2} \theta_{W} \left(1 + 3t^{2}\right)} \left(\frac{M_{Z'}^{6}}{4 M_{Y_{1}}^{2} M_{Y_{2}}^{2}} +  \frac{2M_{Z'}^{4}}{M_{Y_{1}}^{2}} -  \right. \nonumber \\
&& \left.  \frac{M_{Z'}^{4}}{M_{Y_{2}}^{2}} + \frac{3 M_{Z'}^{2} M_{Y_{1}}^{2}}{2 M_{Y_{2}}^{2}}  - \frac{9M_{Z'}^2M_{Y_2}^2}{2M_{Y_1}^2}  -5 M_{Z'}^{2} +  \frac{M_{Y_{1}}^{6}}{4 M_{Z'}^{2} M_{Y_{2}}^{2}} - \frac{M_{Y_{1}}^{4}}{M_{Z'}^{2}} +  \right.  \nonumber \\
&& \left. \frac{3 M_{Y_{1}}^{2} M_{Y_{2}}^{2}}{2  M_{Z'}^{2}} + \frac{M_{Y_{2}}^{6}}{4 M_{Z'}^{2} M_{Y_{1}}^{2}} + \frac{M_{Y_{2}}^{4}}{M_{Z'}^{2}} -  \frac{M_{Y_{1}}^{4}}{M_{Y_{2}}^{2}}  + 4M_{Y_{1}}^{2} + \frac{2 M_{Y_{2}}^{4}}{M_{Y_{1}}^{2}} - 5 M_{Y_{2}}^{2}  \right),
\label{xx}\end{eqnarray}\end{mathletters}
with $h = 1 + 4\tan^2{\theta_W}$. In Eq. (\ref{xx}) $Y_1$ and $Y_2$ are any vector bosons and we 
take for our case $M_{Y_{1}} = M_{Y_{2}} = M_{X}$. Therefore we will have   
\begin{equation}  
\Gamma_{Z' \rightarrow X^{-} X^{+}} =  \frac{\alpha \sqrt{1- 4 M_{X}^2/  s}}{8 M_{Z'} \sin^{2} 
\theta_{W} \left(1 + 3t^{2}\right)}  \left(\frac{M_{Z'}^{6}}{4 M_{X}^{4}} +  
\frac{M_{Z'}^{4}}{M_{X}^{2}} - 8 M_{Z'}^{2} \right).  
\end{equation}\par    
The total cross section for the process $pp \rightarrow qq \rightarrow  P^-P^+$ is related to the 
subprocess $qq \rightarrow P^-P^+$ total cross  section $\hat{\sigma}$, through   
\begin{equation}    
\sigma = \int_{\tau_{min}}^{1} \int_{\ln{\sqrt{\tau_{min}}}}^{-\ln{\sqrt{\tau_{min}}}} d\tau dy  
q\left(\sqrt{\tau}e^y, Q^2\right) q\left(\sqrt{\tau}e^{-y}, Q^2\right)   \hat{\sigma}\left(\tau, 
s\right),   
\end{equation}\noindent 
where $\tau = \left(\tau_{min} = 4 M_P^2/s\right)\hat{s}/s$ and  $q\left(x,Q^2\right)$ is the 
quark structure function.\par

Another form to produce a pair of heavy-leptons is {\it via} the gluon-gluon  fusion, namely through the  reaction of the type $pp \rightarrow g g   \rightarrow P^-P^+$. Since the final state is neutral, the $s$ channel involves  the exchange of the two neutral Higgs boson $H_{3}$ $(h)$ and $Z'$.  The exchange of a photon is not allowed by C conservation (Furry's theorem), which also indicates that only the axial-vector couplings of the neutral gauge boson $Z$ $\left(Z'\right)$ contribute to this process. Moreover the axial-vector coupling of the $Z$ to the exotic lepton is equal to zero. Therefore there is not contribution of this particle to the total cross section.  The interference of the $Z$ with the $H_{3}$ and $h$ vanishes, since the first one is antisymmetric in the gluon polarization, while the latter are symmetric. Therefore, the differential cross section for this reaction can be expressed as
\begin{equation}  
\frac{{d\hat\sigma}}{d \Omega} = \frac{1}{64 \pi^{2} \hat{s}} \left(\overline{\left|M_{Z'} \right|^2} + \overline{\left|M_{H_3}\right|^{2}} + \overline{\left|M_{h} \right|^{2}} + 2 {\it Re} \overline{M_{H_3}M_h}\right).
\end{equation}   
Writing separately the cross-section for the subprocesses, we will have first for the $Z'$
\begin{equation}  
\hat{\sigma}_{Z}^q = \frac{\left(g_{A}^{\prime PP}\right)^{2} \alpha^2  \alpha_s^2}{512\pi\sin^4\theta_W \cos^{4}\theta_W} \frac{M_P^2}{M_Z'^4} \beta  \left|\sum_{m_q = m_u,m_d,m_{J}} g_{A}^{q}  \left( 1 + 2 \lambda_q I_q  \right) \right|^2, \label{sqz}\end{equation}
where the summations run over all generations and $\lambda_q = m_q^2/\hat{s}$. For the $H_{3}$ we have
\begin{eqnarray}  
\hat{\sigma}_{H_{3}}^{q} = &&\frac{\alpha_{s}^{2} \hat{s} (4 \hat{s} - 5 M_{P}^{2} )}{32768 \pi^{3}} \frac{M_P^2}{v_\chi^4} \beta  \left|\chi_3 (\hat{s}) \sum_{q=J} 2 \lambda_q +  \lambda_q (4 \lambda_q -1) I_q  \right|^2, 
\label{qh} \end{eqnarray} 
where the summations run only over $J_{1}, J_{2}$ and $J_{3}$. We also define 
\begin{equation}
\chi_i (\hat{s}) = \frac{1}{\hat{s} - M_{H_{i}}^{2} + i M_{H_{i}} \Gamma_{H_{i}}}
\end{equation}
with $\Gamma_{H_i}$ being the Higgs-boson total width. The $H_3 \rightarrow J \bar{J}$ decay widths are
\begin{equation}
\Gamma\left(H_3 \to J\overline{J}\right) = \frac{\sqrt{M^2_{H_3} - 4M^2_{J}}M^2_J}{16\pi M_{H_{3}}^{2} v_\chi^2}\left(2M^2_{H_3} - 5M^2_J\right).
\label{jj}\end{equation}
On the other hand the $H_{3} \rightarrow P \bar{P}$ decay widths are
\begin{equation}
\Gamma\left(H_3 \to P\overline{P}\right) = \frac{\sqrt{M^2_{H_3} - 4M^2_P}M^2_P}{64\pi M_{H_{3}}^{2} v_\chi^2}\left(4M^2_{H_3} - 5M^2_P\right).
\label{pp}\end{equation}
We notice here that the cross section $\hat\sigma^q_h$ and the decays $h \to J\bar J$ and $h \to P\bar P$ give contributions analogous to (\ref{qh}), (\ref{jj}) and (\ref{pp}), respectively with $M_{H_3}$ replaced to $M_h$.\par
The loop integrals involved in the evaluation of the elementary cross section can be  expressed in terms of the function $I_i\left(\lambda_i\right) \equiv I_i$ which is defined through   
\begin{equation} 
I_i = \int_0^1\frac{dx}{x}\ln\left[1 - \frac{\left(1-x\right)x}{\lambda_i}\right] =  \left\{\begin{array}{rr}  - 2 \left[\sin^{-1}\left(\frac{1}{2\sqrt{\lambda_i}}\right)\right]^2, & \quad \lambda_i > 1/4, \cr \frac{1}{2} \ln^2\left(r_+/r_-\right) + i\pi\ln\left(r_+/r_-\right) - \pi^2/2, & \quad \lambda_i < 1/4\end{array}\right., 
\end{equation}
with $r_\pm = 1 \pm \sqrt{1 - 4 \lambda_i}$. Here, $i = q$ stands for the particle (quark) running in  the loop.\par   
The diagram with a $Z'$ in the $s$ channel does not exhibit the resonance for $\hat{s} = M_{Z'}^{2}$, since the on shell production of a massive   spin-one particle on shell from two massless spin-one particle is  forbidden (Yang's theorem) \cite{yang}. The total cross section for the process $pp \rightarrow gg \rightarrow  P^-P^+$ is related to the subprocess $gg \rightarrow P^- P^+$ total cross section $\hat{\sigma}$ through  
\begin{equation}   
\sigma = \int_{\tau_{min}}^{1}  \int_{\ln{\sqrt{\tau_{min}}}}^{-\ln{\sqrt{\tau_{min}}}} d\tau dy  G\left(\sqrt{\tau}e^y, Q^2\right) G\left(\sqrt{\tau}e^{-y}, Q^2\right)   \hat{\sigma}\left(\tau, s\right), 
\end{equation}
where $G\left(x,Q^2\right)$ is the gluon structure function.

As an example for the decay of the heavy exotic leptons we have $T^{-} \rightarrow E^{-} \bar{{\nu}_{e}} \nu_{\tau}$, {\it i. e.}, 
\begin{equation}
\Gamma (T^{-} \rightarrow E^{-} \bar{{\nu}_{e}} \nu_{\tau} ) = \frac{G^{2} M_{T}^{5} M_{W}^{4}}{192 \pi^{3} M_{V}^{4}} \left(1- 8 \delta + 8 \delta^{2} - \delta^{4} - 12\delta^{2} \ln \ \delta \right),
\end{equation}
where $\delta = M_{E}^{2}/M_{T}^{2}$ and $M_{V}$ is the mass of the new single charged gauge boson which can run from 200 GeV to 3000 GeV \cite{JJ97}.

\section{Results and conclusions}    

The process for the heavy-lepton production in hadronic colliders was well studied in the literature and was shown that the dominant  contribution are the well  known Drell-Yan process 
(Fig. 1a) and gluon-gluon fusion (Fig. 1b) \cite{cie1,dic}.\par
We present the cross section for the process $pp \rightarrow P^{-}P^{+}$, involving the Drell-Yan mechanism and the gluon-gluon fusion, to produce the 3-3-1 heavy-leptons. In Fig. 5 we show the different cross sections for different masses of neutral boson  $Z'$, $H_3$ and $h$. We see that the 
Drell-Yan and the gluon-gluon fusion production are suppressed for $2M_{P} > M_{Z^\prime} (M_{H_3}, M_{h})$ where the $Z^\prime$, the $H_{3}$ and $h$ later must produce the heavy-lepton pairs. This figure still exhibit the resonance peaks associated with the boson $Z'$ and the two neutral Higgs bosons. \begin{figure}[h]
\begin{center}  
\epsfxsize=6in  
\epsfysize=7 true cm  
\centerline{\epsfbox{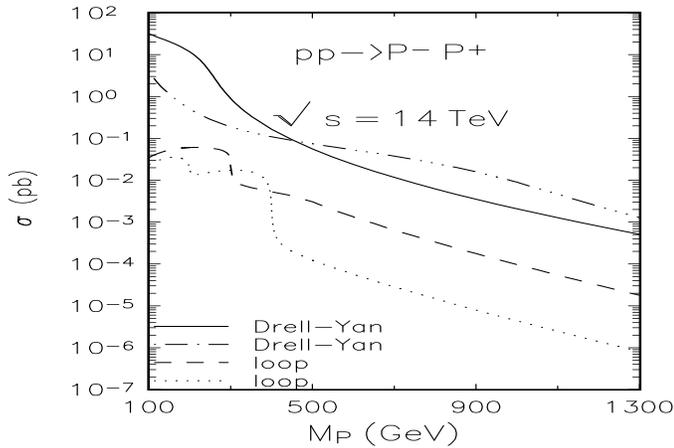}}  
\end{center}  
\caption[]{Total cross section for the process $pp \rightarrow P^-P^+$ as a 
function of $M_{P}$ at $\sqrt{s} = 14$ TeV using the mechanism of the Drell-Yan and of the gluon-gluon fusion. For the Drell-Yan case we take the mass of the boson $Z'$ equal to $0.6$ TeV (solid line) and 2 TeV (dot dot dashed line) and for the gluon-gluon fusion we have for the mass of the boson $Z'$  equal to $0.6$ TeV (med dashed line) and 2  TeV (dotted line).}
\label{fig:5}\end{figure}
Considering that the expected integrated luminosity for the LHC will be of order of $10^{5}$ pb$^{-1}$/yr and taking the mass of the heavy-lepton equal to 300 GeV, it could be seen in the LHC, that the  Drell-Yan production dominates over gluon-gluon fusion. 
\begin{figure}[h]
\begin{center}  
\epsfxsize=4in  
\epsfysize=7 true cm  
\centerline{\epsfbox{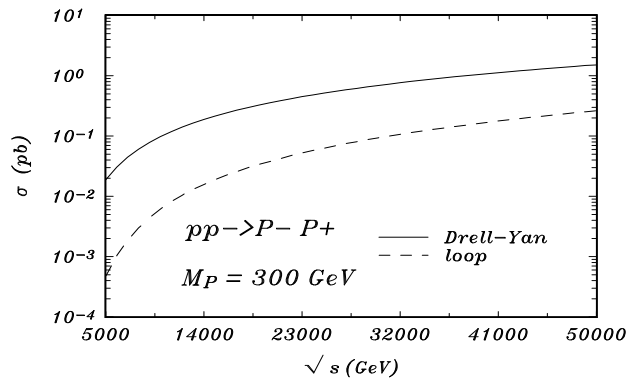}}  
\end{center}  
\caption[]{Total cross section for the process $pp \rightarrow P^-P^+$ {\it versus} $\sqrt{s}$ in according with the mechanism of the Drell-Yan (solid line) and the gluon-gluon fusion (dashed line). Here was taken for the mass of the heavy exotic leptons equal to 600 GeV.}
\label{fig:6}\end{figure}
So for the Drell-Yan mechanism we have a total of $\simeq 8.8 \times 10^4$ and $1.7 \times 10^4$ heavy-leptons produced per year if we take the mass of the boson $Z'$ equal to $0.5$ TeV and 2 TeV. For both figure was taken, $M_{J_1} = 200$ GeV, $M_{J_2} = 300$ GeV, $M_{J_3} = 500$ GeV and for the mass boson $M_V = 800$ GeV. For the gluon-gluon fusion we will have a total of $\simeq 2 \times 10^3$ and $1.8 \times 10^3$ heavy-leptons produced per year if we take the masses of the Higgs $H_{3}$ $(h)$ equal to 600 (600) GeV and 400 (800) GeV, respectively. Here we take for the mass of the boson $Z'$ and the quarks the same values as for the Drell-Yan case. In addition we take for $v_\chi = 700$ GeV.\par
\begin{figure}[h]
\begin{center}  
\epsfxsize=6in  
\epsfysize=7 true cm  
\centerline{\epsfbox{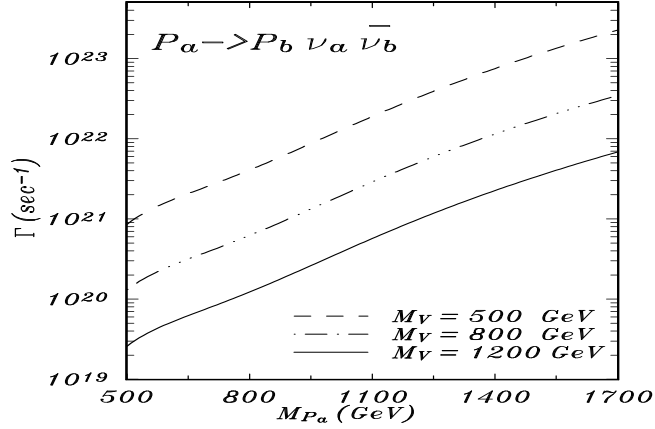}}  
\end{center}  
\caption[]{The partial decay widths of the heavy-lepton $T$ {\it versus} $M_{T}$ for several masses of the new single charged gauge boson $V$: For $M_{V} = 0.5$ TeV (dashed line), $M_{V} = 0.8$ TeV (dot dot dashed line) and $M_{V} = 1.2$ TeV (dotted line).}
\label{fig:7}\end{figure}
From Eq. (\ref{sqz}) it is to see that the cross-section {\it via} gluon-gluon fusion is very sensitive to the vacuum expectation value, so if we take $v_\chi = 500$ GeV and the same values as was taken above for the mass of the boson $M_{Z^{'}} = 2000$ GeV, we will have a total of $3.5 \times 10^{3}$ heavy leptons produced per year, that is nearly the value of $1.7 \times 10^4$, that we have received for the case of Drell-Yan mechanism.\par
\begin{figure}[h]
\begin{center}  
\epsfxsize=6in  
\epsfysize=7 true cm  
\centerline{\epsfbox{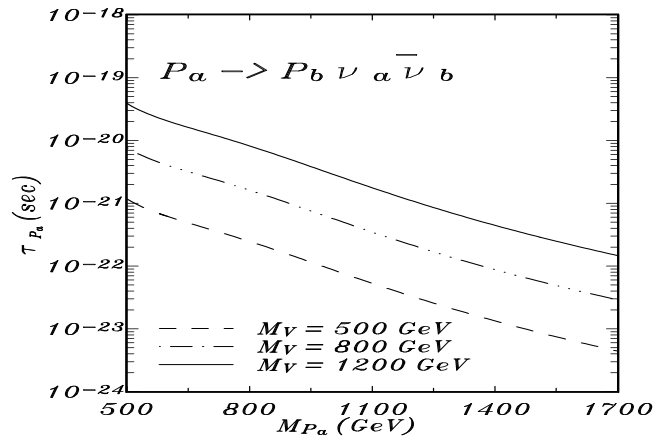}}  
\end{center}  
\caption[]{Lifetime of the heavy-lepton $T$ vesus the $M_{T}$ for several masses of the new gauge boson $V$: $M_{V} = 0.5$ TeV (dashed line), $M_{V} = 0.8$ TeV 
(dot dot dashed line) and $M_{V} = 1.2$ TeV (dotted line).}
\label{fig:8}\end{figure}
The total cross sections for exotic heavy-leptons produced by the mechanisms of Drell-Yan and of gluon-gluon fusion {\it versus} $\sqrt{s}$ are showed in Fig. 6. Here was taken for the mass of the heavy exotic leptons  $M_P = 300$ GeV, for the Higgs masses $M_{H_{3}} = 400$,  $M_{h} = 800$ GeV (for the other parameters was taken the same values as in Fig. 5). From this figure it is to see that for very high energy we will have for every heavy lepton produced {\it via} gluon-gluon fusion, five or six leptons produced {\it via} Drell-Yan. \par
The Fig. 7 shows the partial width {\it versus} the mass of the decaying particle for the process $T^{-} \rightarrow E^{-} \ \bar{\nu_{e}} \ \nu_{\tau}$, considering that the mass of the boson $V$ runs from $500$ GeV to $1200$ GeV. The Fig. 8 shows the $T$-lepton lifetime {\it versus} $M_{T}$. This figure exhibits that the heavy-lepton can be short-lived for the masses of the boson $V$ here considered.\par
The main background for the signal, $q \bar{q} \rightarrow P^{-}  P^{+} \rightarrow \bar{\nu} \bar{u} J_{1} \ (\nu u \bar{J_{1}} )$, could comes from the process $q \bar{q'} \rightarrow W^{-} W^{+} (Z \ Z) \rightarrow q \bar{q'} q \bar{q'} (q \bar{q} q \bar{q} )$, but this background can be eliminated by measuring the undetected neutrino contribution $p_{\nu T}$, since all hadrons with appreciable $p_{T}$ are detected. Then the overall $p_{T}$ imbalance for detected particles gives this undetected $p_{\nu T}$. We could have another backgrounds such as $q \bar{q'} \rightarrow W^{-} W^{+} (Z \ Z) \rightarrow e \nu q \bar{q'} (e^{+} e^{-} q \bar{q} )$, but this backgrounds also can be eliminated, then the number of jets are different. The very striking signal comes from the reaction $P_{a}^{+} \rightarrow \ell^{-} \ell^{+} P_{b}^{+}$ since it consists of three charged particles.\par
In summary, we showed in this work that in the context of the 3-3-1 model the signatures for heavy-leptons can be significant in LHC collider. Our study indicates the possibility of obtaining a clear signal of these new particles with a satisfactory number of events.

\acknowledgements 

One of us (M. D. T.) would like to thank the Instituto de F\'\i sica Te\'orica of the UNESP, for the use of its facilities and the Funda\c c\~ao de  Amparo \`a Pesquisa no Estado de S\~ao Paulo (Processo No. 99/07956-3) for full financial support. We are grateful to Prof. O. J. P. \'Eboli for calling the attention to some points.

\end{document}